\documentclass[8pt]{article}
\usepackage{amsmath}
\usepackage{amsthm}
\usepackage{marginnote}
\usepackage[disable]{todonotes}
\usepackage{mathtools}
\usepackage{float}
\usepackage{tcolorbox}  % Boxes
\usepackage{booktabs} % Tables
\usepackage{caption}
\usepackage{tabularx}
\usepackage{multicol}
\usepackage{enumitem}
\usepackage{lineno}
\usepackage{inputenx}

\usepackage[
backend=biber,
style=numeric,
citestyle=numeric
]{biblatex}

 \usepackage{graphicx}
 \usepackage[font={small}]{caption}
 \usepackage{authblk}
\setuptodonotes{color=green!10}

\theoremstyle{definition}  % Make all theorems non-itallic
\newtheorem{theorem}{Result}

\bibliography{bibliography}

\title{Revisiting the relevance of traditional genres: a network analysis of fiction readers' preferences}
\author{Taom Sakal, Stephen Proulx}
\affil{\small Department of Ecology, Evolution and Marine Biology, University of California, Santa Barbara, USA}
\date{}

\begin{document}

	%\todototoc
	%\listoftodos
	\maketitle

\begin{abstract}
    We investigate how well traditional fiction genres like Fantasy, Thriller, and Literature represent readers' preferences. Using user data from Goodreads we construct a book network where two books are strongly linked if the same people tend to read or enjoy them both. We then partition this network into communities of similar books and assign each a list of subjects from The Open Library to serve as a proxy for traditional genres. Our analysis reveals that the network communities correspond to existing combinations of traditional genres, but that the exact communities differ depending on whether we consider books that people read or books that people enjoy.

    In addition, we apply principal component analysis to the data and find that the variance in the book communities is best explained by two factors: the maturity/childishness and realism/fantastical nature of the books. We propose using this maturity-realism plane as a coarse classification tool for stories.

\end{abstract}

\section{Introduction}

% ------- History of Storytelling -------
Storytelling is one of our oldest arts. It dates to prehistory and has shaped who we are as individuals and as societies. Story itself comes in diverse forms: oral traditions, art, music, writing, plays, movies, and even casual conversation. Storytelling is a human universal \cite{brownHumanUniversalsHuman2004} and some argue that it evolved as a way to coordinate group behavior by communicating the expectations, norms, and rules of a society \cite{smithCooperationEvolutionHuntergatherer2017}. Some even argue that storytelling is what makes us human \cite{gottschall2012storytelling}.

% ------------- Storytelling makes money -------------
Whether or not it makes us human, it certainly makes us money. Fiction in the form of novels and movies both represent multi-billion dollar industries. The Association of American Publishers reported a 2017 revenue of approximately \$4.38 billion for adult fiction book sales \cite{andersonNewStatistics20172018}. The US book industry as a whole is currently worth \$29.4 billion \cite{ibisworldIBISWorldUSIndustry} and has a total 300 million titles in print with 100,000 new ones published a year. About 2.7 billion books are sold each year \cite{yucesoySuccessBooksBig2018a}.

% -------------- Social value of books -------------------
Stories in written form also have enormous historical, cultural and social value. The modern English novel appeared no later than the 18th century \cite{wattRiseNovel2001}, and since then pleasure reading has been associated with

\begin{itemize}
	\item Increased general knowledge.   \cite{cunninghamWhatReadingDoes1998} \\
	\item Insight into other other cultures \cite[Meek 1991, as cited in][]{clarkReadingPleasureResearch2006}\\
	\item Understanding of human nature and decision making. \cite[][Bruner 1996, as cited in]{clarkReadingPleasureResearch2006}
	\item Reduced feelings of loneliness in adults. \cite{rane-szostakPleasureReadingOther1995}
	\item Better vocabulary and verbal reasoning.
\end{itemize}

The last point comes with the surprising statistic that the average book written for preschoolers contains a higher frequency of rare words than the average conversation between college graduates. This is a byproduct of print containing more rare words than everyday speech  \cite{cunninghamWhatReadingDoes1998}.

% This paragraph is not essential to the narrative but so striking it might be worth leaving in, if only to help the reader remember our paper

% ---------------------------------- Introducing Elemental Genre  --------------------------------------------------

Generally fiction is organized into genres like Fantasy, Horror, and Literary which we call \textit{traditional genres}. They capture a combination of setting, plot, and tropes that define groups of stories. While the groups often blur together, traditional genre forms a rough way to describe a reader's tastes. In America the most popular traditional genres are Mystery, Thriller, and Crime \cite{yucesoySuccessBooksBig2018a}.

But do traditional genres describe reader preferences? A group of professional fiction have argued that they do not  \cite[ep 11.1]{sandersonWritingExcusesSeason}. Instead they suggest that \textit{elemental-genres}, which group stories by the emotion they make the reader feel, are what determines preference. While elemental-genres are influenced by setting, plot, and tropes, they are not defined by them. In their words,

\begin{quote}
...these elemental genre are the things that make you read. The things that make you turn the page. The emotional resonance you are shooting for as a writer in your books for the readers to have. The type of element, if you will, that is driving the story forward.
\end{quote}

They further suggest that a major reason people read is to experience certain emotions. Stories are then tools with which people can ``hack" their brain and experience their chosen emotions. This suggestion is plausible. \cite{yuPracticalTypologyAdult1999} and \cite{moyerAdultFictionReading2005} showed that many people consistently read as a way to escape the real world.  Viewed this way, elemental-genres classifying books by the emotion they elicit. 

%It is not difficult to imagine that this generalizes to books as a tool to elicit certain emotions. Romance, erotica, and power fantasies come to mind as examples of this.

% TODO This is the gist of what these papers say but not exactly. There's also more in them that support this line of reasoning which I don't have time to re-read right now.

%--- Table -------------
\begin{table}[H]
	
 	\begin{center}
 		 \scalebox{.7}{
 		\begin{tabular}{ p{2cm}  p{4cm} p{4cm}  p{5cm} }

 		\toprule
 			\textbf{Elemental Genre} & \textbf{Driving element} & \textbf{Example} & \textbf{Reason for Examples}\\
 			\midrule

 			Wonder & A sense of wonder and awe & Harry Potter (1st book),  Ringworld & Harry's amazement at all the details of the wizarding world; awe at the sheer scales involved in Ringworld.\\

 			Idea & Stories with ``what if'' questions that fascinate the reader  & 1984, Holes & Asks what if governments could rewrite history on their whim or if there existed a desert camp where children had to dig holes for a mysterious purpose?  \\

 			Adventure &  Stories that focus on characters pushing their limits, the thrill of doing something.amazing. & Indiana Jones, Jackie Chan movies  & About if Indiana Jones can get cool objects, about Jackie Chan doing amazing martial arts.\\

 			Horror & Sense of fear, dread, and powerlessness, often towards something that already happened. & Stephen King Novels & Follows a character who will suffer horribly, even if they survive.\\

 			Mystery &  Stories that present a puzzle. They are read for the thrill of solving the puzzle & Sherlock Holmes & From slowly revealed clues the reader and characters must figure out who committed a crime. \\

 			Thriller & Constant, fast paced anxiety. Fear and dread of the known.  & The Da Vinci Code & Has short chapters that end in cliffhangers and constant tension. \\

 			Humor & Stories that are driven by their humor. Often read for the funny twists and punchlines that re-contextualize everything  & The Hitchhiker’s Guide to the Galaxy & Constantly sets up situations then re-contextualizes. For example,  “the ships hung in the sky, in much the same way that bricks don’t.”\\

 			Relationship & How a relationship between two characters develop & Romance Novels &[todo: it's important to emphasize that these stretch beyond romance. Add other example.] \\

 			Drama & A character's journey, how it transforms them, and how that effects all those around them. & King Lear & King Lear descends into madness and this affects him and all the characters he knows.\\

 			Issue & Characters dealing with a social issue or conflict (ex racism).  & A Christmas Carol & Scrooge must come to terms with his greed.\\

 			Ensemble & Many different specialists interacting and working together to achieve a goal & Firefly, Star Trek, most heist novels & Each follow a team of specialists \\
 			\bottomrule
 		\end{tabular}
 	}
 		\captionof{table}{The elemental-genres and given examples according to  \cite{sandersonWritingExcusesSeason}. They do not claim their list is canonical, rather just one of many possible lists.}
 	\end{center}

\end{table}

  \begin{table}[H]
  	\centering
  	\resizebox{\textwidth}{!}{%
  		\begin{tabular}{@{}llll@{}}
  			\toprule
  			\multicolumn{4}{c}{traditional genres}                                                                   \\ \midrule
  			Action Adventure     & Fantasy Romance                  & Law Enforcement      & Science Fiction         \\
  			Amateur Sleuth       & Gothic                           & Legal Thriller       & Science Fiction/Fantasy \\
  			Category Romance     & Graphic Novel                    & Literary             & Space Opera             \\
  			Contemporary Romance & Hard-Boiled                      & Medical Thriller     & Thriller                \\
  			Cozy Mystery         & Heroic Fantasy/Sword and Sorcery & Mystery              & Traditional British     \\
  			Dark Fantasy         & Historical                       & Paranormal Romance   & Traditional Regency     \\
  			Epic Fantasy         & Historical Mystery               & Political Thriller   & Urban Fantasy           \\
  			Espionage/Spies/CIA  & Historical Romance               & Private Investigator &                         \\
  			Fantasy              & Horror                           & Romantic Suspense    &                         \\
  			&                                  &                      &                         \\ \toprule
  			\multicolumn{4}{c}{Universal Genres}                                                                     \\ \midrule
  			Action               & Horror                           & Thriller             &                         \\
  			Comedy               & Mystery                          &                      &                         \\
  			Drama                & Romance                          &                      &                         \\
  			&                                  &                      &                         \\ \bottomrule
  		\end{tabular}%
  	}
  	\caption{A sample list of traditional genres. These vary greatly depending on the source, and can number from a handful to hundreds of genres. The first set is taken from the subgenre list of FictionDB.com. These represent a more specific style towards classification. A counterpoint to that is the broad universal-genres list taken from TvTropes.com.}
  	\label{tab:elemental-genres}
  \end{table}

% --------- Elemental Genre Example Example --------------------

%For example, say that a reader likes Harry Potter. If readers tend to stay in traditional genres then we might expect this reader to also enjoy Game of Thrones. After all, it has all of the fantasy tropes: dragons and adventure and magic. \todo{Need better example.}

%Yet we would not expect a Harry Potter fan to enjoy Game of Thrones because they have totally different tones. Harry Potter is filled with a sense of wonder and awe -- every corner of the world is painted with small details about how wizards and witches lives their lives. Contrast this with the scheming and politics of Game of Thrones. We might rather expect a fan of Harry Potter to enjoy a book like Ringworld more \todo{See what examples Sanderson gave}. Even though this book is firmly in the traditional genre of hard science fiction it is filled with wonder and awe, only this time about a planet in shape of a ring rather than a world of magic.

% -------- Elemental genre as example of alternative classification -------------------

This line of thought raises the possibility that traditional genre poorly describe reader's tastes. This would mean our default classification system is misaligned, and does not reflect the underlying reasons people read enjoy certain stories. And classification system matters to readers. Almost one fifth of readers search for books by genre \cite{sear1986readers}, and in one case study a library increased their circulation by thirty six percent by their fiction and some of their nonfiction down by category \cite{shearer1996guiding}. Better classification systems may improve these two metrics, and help readers discover stories they might not have otherwise.

% TODO: read these two citations in more detail. The second one we only need to look at the chapter on genre.

% CUT: By a more accurate classification system  Such a system should reflect reader's preferences. But there are two ways to interpret preference: what books people choose to read, and what books people people enjoy. A bookseller may be tempted to only study which books readers prefer to read (and hence buy), but this does not account for user reviews and word of mouth which heavily influence sales \cite{chevalierEffectWordMouth2006a}. Conversely, an author may be tempted to only consider the books people enjoy,
%MARK 5/19/20
%These form two separate but related measures of preference and we study both.

% --------------------------------------------------

Classification systems also matter to writers. Established authors, when writing a new work, must balance the familiar and unfamiliar. They must keep elements that their fans enjoyed without keeping so many that their new word feels like a rehash of their previous works  \cite{cardElementsFictionWriting2010}. Staying within traditional genre does this through keeping plot, setting, or tropes constant. Staying within elemental-genre is an alternative, one that allows the writer to keep the emotional resonances constant while varying  plot, characters, and setting \cite{sandersonWritingExcuses112016}.

In this study we create a new classification system by incorporating reader preference data downloaded from Goodreads. From these we can create two new emergent forms of genre: reader genre and enjoyment genre. The first gives groupings of books that are often read together by the same person. The second one gives books that are often enjoyed by the same person. For clarity we will call these ``communities'' instead of genres, as they stem from communities within a book network we create. 

We then compare these communities to those given by traditional genre. As a proxy for traditional genre we gather a list of subjects for each book from The Open Library. To compare we look at the distribution of subjects in each community and see if it differs from the distribution in the network as a whole. We also apply a PCA analysis to find the combinations of traditional genres that best describe our reader and enjoyment communities.

While there has been previous work in automatic classification of books, this mostly consists of classifying the type of book (Fiction, graphic novel, textbook, legal documents, etc). This data driven community and genre style breakdown is far less common and one of the main contributions of our study.

%The following subsections expand on genre and classification systems. %The ideas given here apply to any art form, and we will often borrow examples from well known movies. We also analyses non-fiction works here,

%================================================
\subsection{Defining Genre.}
%================================================
The idea of classifying books into genres have been around since ancient times. Aristotle gives us one of the first classifications which include the genres of tragedy, comedy, and the epic \cite{butcherPoetics2006}. But starting around 1820 literacy increased and novels moved from being aimed at general audiences to specific readers. This created something like the genre we think of today -- a way to describe reader's taste \cite{morettiGraphsMapsTrees}.

% Genre is a classification scheme
But just as the list of genres change depending on who you ask, so too does the definition of genre. The dictionary defines genre as “a category of artistic composition, as in music or literature, characterized by similarities in form, style, or subject matter” \cite{websterdictionaryWebsterDictionary}.  This is an overloaded definition as it can refer to medium (plays, fiction, non-fiction, graphic novel, etc) along with shared conventions (plot, setting, tropes, etc). 

\cite{ericksonRhymePunishmentCreation1999} gives a more academic definition.

\begin{quote}
	``...a patterning of communication created by a combination of the individual, social and technical forces implicit in a recurring communicative situation. A genre structures communication by creating shared expectations about the form and content of the interaction, thus easing the burden of production and interpretation."
\end{quote}

Part of what makes genre such a strong concept is its inherent vagueness and flexibility to describe different works. The categories themselves are fluid and change with the times. Genres tend to appear and disappear in clusters every 25-30 years due to generational turnover in the reading public \cite{morettiGraphsMapsTrees}, and the details of how this occurs can be modeled computationally in \cite{sackSimulatingCulturalEvolution2013}.

The vagueness and changing nature of genre can make labeling a story difficult.  The famous example here is Star Wars. Is it considered Science Fiction because the setting is futuristic? Or is it Fantasy because there is magic in the form of ``the Force" and the plot line follows the typical fantasy plot? Some one, some say the other. Some say it is in it's own subgenre of Science-Fantasy. Some say it is in both Science Fiction and in Fantasy, but that there is no need to define a sub-genre of Science-Fantasy.

In the fiction community the vague and flexible definition serves well, but for quantitative work we require something more precise. For us ``genre" will refer to any system that classify books into communities.  There already exists a vast literature concerning these for fiction. There are five points that a useful classification scheme must follow, according to \cite{bakerFictionClassificationSchemes1987}.

%...where books within the same community are on average more similar to each other than two books in different communities. The exact groupings of books will depend on how one measures similarity. If similarity is determined by setting, plot, and tropes then the groupings give backs the traditional genres. If we instead measure similarity by emotional tone then we produce elemental genres. If you measure by format of the book you will have genres like “short story,” “novella,” “novel,” and “graphic novel.” % Maybe can remove this paragraph.

% %

% \begin{table}[H]
	
% \begin{center}
%     \resizebox{\textwidth}{!}{%
% 	\begin{tabular}{ l  l }
%     \toprule
% 	 \textbf{Genre/Community Type} & \textbf{Similarity Measure} \\
% 	 \midrule
% 	 Traditional & Books with the same settings and tropes. \\
% 	 Market & Books which are frequently bought together. \\
% 	 Reader &  Books that are frequently read by the same person.\\
% 	 Enjoyment & Books frequently enjoyed by same person.\\
% 	 Elemental &  Books with the same driving emotion.\\
% 	 Format & Books written in the same medium (ex. short stories, comics, novels.)\\
% 	 \bottomrule

% 	\end{tabular}
% 	}
% 	\captionof{table}{Some possible classification schemes.}
% 	\end{center}

\begin{enumerate}
	\item Make it easy to find the books you want.
	\item Have different types of divisions with which one can use to filter their search (e.g. broad category, genre, format, literary quality).
	\item Expose readers to authors/books they would have otherwise overlooked.
\end{enumerate}

There are two more controversial points too.

\begin{enumerate}
	\setcounter{enumi}{3}
	\item Each book must be in only one genre. (This is common in physical libraries or bookstores where a book can only be placed in a single section.)
    \item Each author must be in only one genre. (No splitting of authors.)
\end{enumerate}

The fourth point is the most controversial. Traditional and emotional genre do not obey it. However, the genres we create in this study will because it simplifies the analysis.

%There are infinite possible genre classes but we focus on reader and enjoyment genre. If a reader only reads books with TV and movie adaptations, then that is a reader genre. If a reader only reads books recommended by The New York Times, then that is a reader genre. Reader genre is defined by what the audience reads, and enjoyment defined by what the reader enjoys. For clarity we will call the reader and enjoyment genres ``communities," as we create them from book communities within a network.

%If this elemental-genre idea is true we might expect our enjoyment-communities to overlap with elemental-genres while maybe not overlapping with reader-communities. Likewise we might expect traditional genre to overlap with reader-communities but not enjoyment. If these are true then we have evidence that elemental-genre gets at the core of why readers enjoy certain books, and that setting and plot matter much less. It would also be interesting to see if market-genre overlaps with enjoyment or elemental-genre. This would give us a view into the difference in what people choose to read versus what they enjoy.

%We expect some degree of overlap no matter what. For example, one of the largest industries in fiction is the romance industry. The question is how much of it is from the emotions and how much from the traditional genre. Romances have their own distinct plots and emotions, meaning we expect them to have the same emotional genre. Indeed, romance readers tend to stay in romance.

%=============================
\section{Methods}  % Draft 2
%============================

We take users and their book ratings from Goodreads and construct a similarity network of books. In this network nodes are books and weighted links correspond to how similar two books are under a given similarity measure. The Python notebook file in the supplementary materials walks through the exact process we use.
%\todo{This is in genre breakdown.tex}

\begin{center}
\includegraphics[scale=.25]{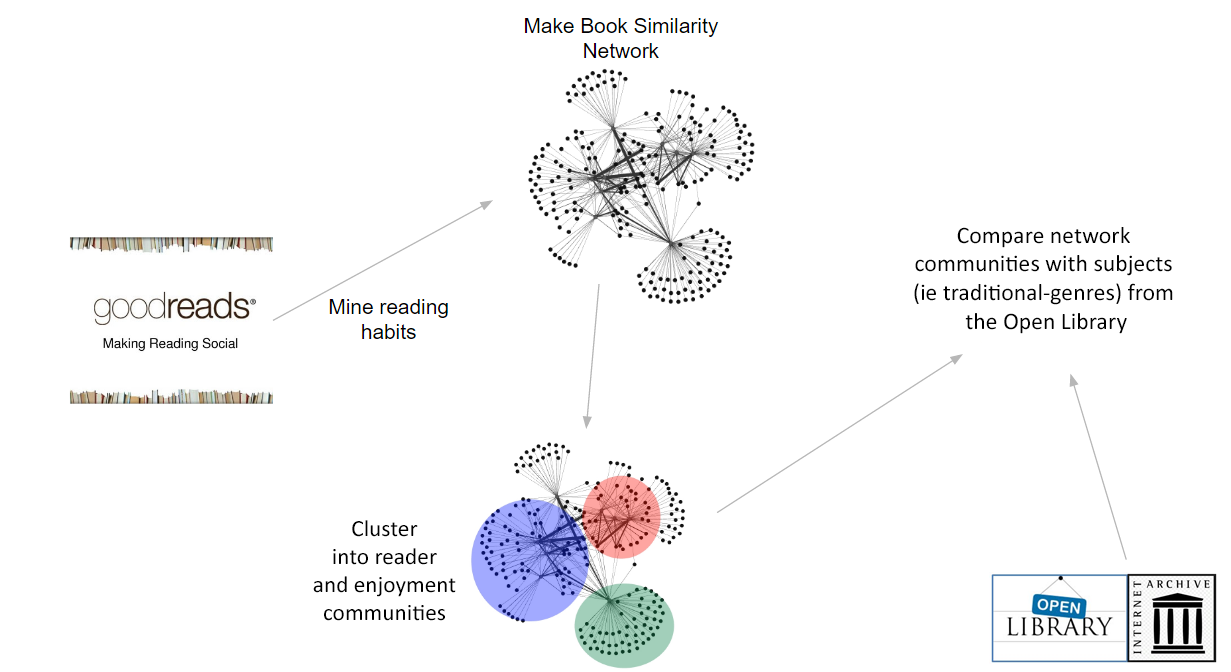}
\end{center}

\subsection{Data}

Our analysis requires four core pieces of data.

		\begin{itemize}
			\item The users.
			\item The books each reader read.
			\item The rating each reader gave those books.
			\item The traditional genre(s) of those books.
		\end{itemize}

The first three items allows us to make the similarity networks for reader genres and enjoyment genres. The last item allows us to make the traditional genres. We gather the first three from Goodreads, a social network where users catalogue, rate, and write reviews for books. The site has approximately 65 million user accounts and functions as a place to track, rate, and discover new books. Goodreads provides customized recommendations and friends on the platform can directly suggest books.

With the permission of Goodreads we gathered data from 99,909 randomly chosen users. Note that Goodreads caps us to downloading only the two hundred most recent books of a user, so the few super-readers in our dataset only have their more recent two hundred books. We removed any user with only a single book as they give no information in our projection. This left us with 26,076 users, or twenty six percent of the original amount. Out of these each user has a list of books that they have read, along with their ratings of the book. We filtered any book with less ten reads for computability purposes. These books tended to be little known books with less subject data.

Note that we are unable to share the raw Goodreads data due to legal reasons. The data has also recently become difficult to collect as the Goodreads depreciated their API. One can still collect data via a web scraping. The UCSD Bookgraph project has existing datasets and web scraping code \cite{DBLP:conf/recsys/WanM18,DBLP:conf/acl/WanMNM19}.

For traditional genres we looked each book up by ISBN in The Open Library API and took the list of subjects they had assigned to the book. The Open Library is an an accredited California State library run by The Internet Archive, and has the goal to list every book in existence. The data here comes from a combination of users, libraries, and publishers. A single book can have multiple subjects, from general ones like Psychological Fiction or Love Stories, to specific characters like ``Albus Dumbledore", to platform specific subjects such as ``Staff Picks."  In total we see about six hundred unique subjects, most of which are hyper specific characters or places. To remove this we ignore any subject that appears in a dozen or less books. This leaves 279 subjects, each of which are fairly broad.

%-------------------

\subsection{Creating the Similarity Network}

We create two networks of books which we call the reader network and the enjoyment network. Each book is linked by an edge whose weight is determined by how much the reader-base or fan-base overlaps. To create these networks we begin with a bipartite network of readers and books. The reader nodes represent users of Goodreads and the book nodes represent books. Each reader is linked to books they've read, and links are weighted with what they rated the books. 

\begin{center}
\begin{figure}
\includegraphics[scale=.05]{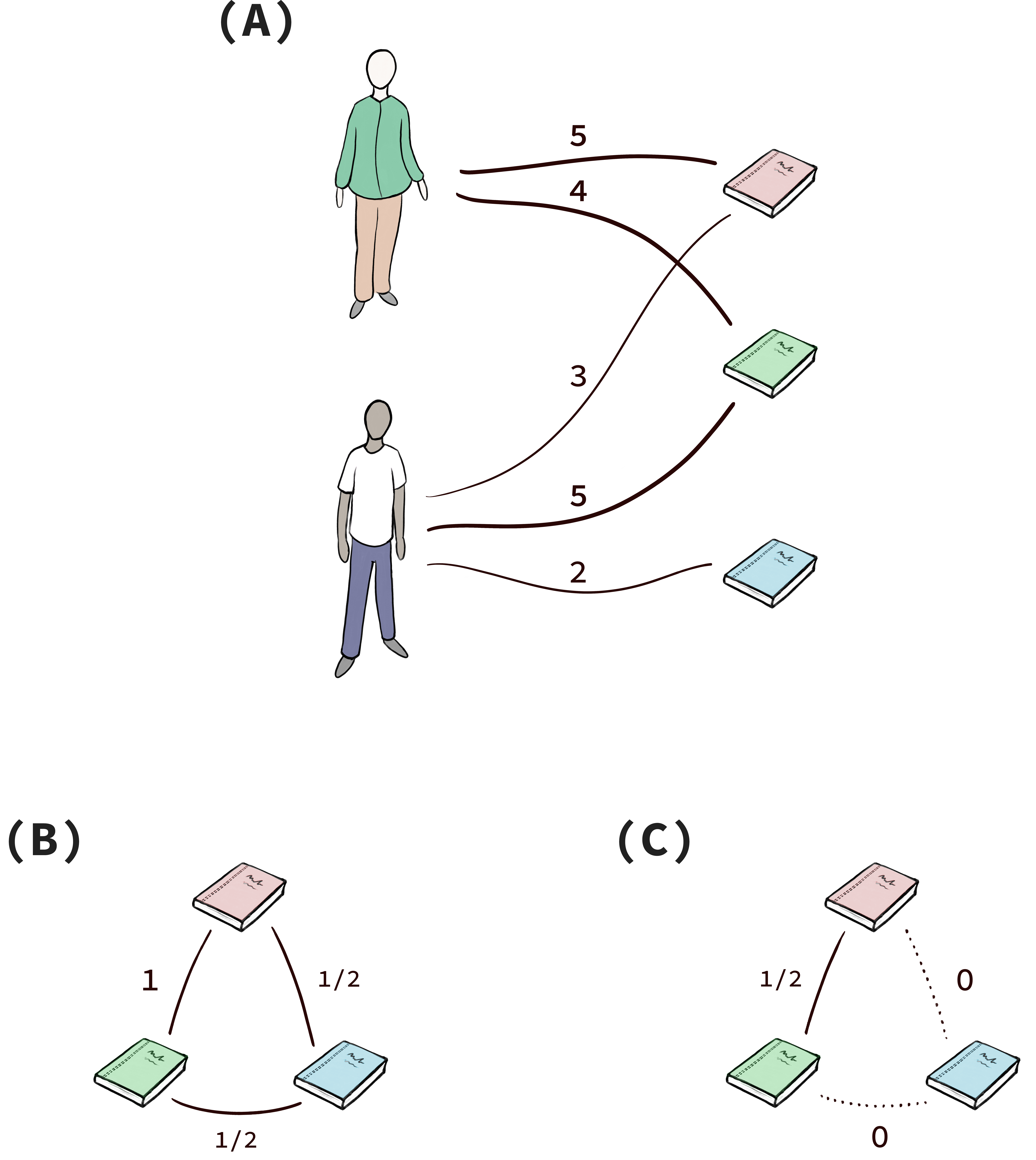}
\caption{(A) An example bipartite network of two readers and three books. The weights are the reader's ratings of each book. (B) The corresponding reader network, made by projecting the bipartite network down onto only the book nodes using the Jaccard Index. (C) The corresponding enjoyment network. This network is formed the same way as the reader network, except we ignore all edges with rating less than four in the original bipartite network. }
\end{figure}
\end{center}
From this we would like project the bipartite network down into a similarity network of only book nodes. For the reader network we calculate the similarity of each book by counting how many readers they share. In this network we ignore weights as we care only about what is read, not what is enjoyed. Two books are similar if many people who read one book also read the other. That is, the reader genre network links two books if they share readers, and the weight of the link depends on how much the readers overlap. This is formalized by the Jaccard Index.
\[ w_{i,j} =  \frac{|R_i \cap R_j|}{|R_i \cup R_j|}\]
Here $w_{i,j}$ is the weight of the link between book $i$ and $j$. The set of all people who have read book $i$ is $R_i$ and the set of all people who have read book $j$ is $R_j$. With this similarity measure we can form the reader network by calculating the similarity between every pair of books.

To make the enjoyment genre network we restrict our view from shared readers to shared fanbases. We do this by filtering away all bipartite network links with rating less than four then project down in the same way as before. To justify this method consider how each user contributes to the similarity of two books.

        \begin{enumerate}
        \item If a user rates $book_1$ high and $book_2$ high, then the books have higher similarity.
        \item If a user rates $book_1$ high and $book_2$ low, then the books have lower similarity.
		 \item If a user rates both $book_1$ and $book_2$ low then they have lower similarity.
        \end{enumerate}

These properties reflect our underlying assumptions about how books and people behave: people only have limited time so they want to read the books they like the most. And what people like are a certain subset of tropes or combination of tropes. Most books do not have these elements, so if a user chooses a random book they likely will not enjoy it relative to a book they chose themselves.

This is the rational of why two high ratings mean the books are likely similar -- the reader likely enjoyed both because they shared the same underlying tropes the reader enjoys. Conversely, if a reader dislikes two books it probably is not because they hated the books for the same reason (ex. they both contained dragons and the reader hates dragons) but more likely because they lacked tropes the reader enjoyed.

Why do we choose 4 or more stars to count as enjoying the book rather than 3 stars? Because the average rating for a book among our users is $4.06$ out of $5$. This seems high, but it is typical for online reviews to skew positive. In 2014 Amazon's (who owns Goodreads) average ranking per book was also $4.06$ and Barnes \& Noble's was $4.43$ \cite{chevalierEffectWordMouth2006a}. Because of this we say 4 or 5 stars means the reader enjoyed the book and 3 or fewer corresponds to not enjoying the book. This filters out any rating with less than 4 stars, ensuring that users who did not enjoy a book are not included in that books set of fans. In this case the Jaccard Index measures the overlap in fanbases. After filtering the average rating now becomes 4.56. Here people with no edges are not removed from the dataset but kept in to represent users that read but did not enjoy books.

Finally, before projecting the graph we filter out rare books with less than 10 readers. This is done for computability purposes. Most books in our dataset are read rarely, averaging a mere $3.38$ reads, $2.07$ of which were enjoyable. After this filtering the reader network leaves only the most popular $3.8$ percent of books, yet these popular books account for $55.1$ percent of the links. The story repeats for the enjoyment network: the remaining $2.4$ percent of the most popular books account for $54.2$ percent of the links.

\subsection{Finding and comparing communities }

Next we partition the book networks into communities. We export our graph to Gephi \cite{ICWSM09154} where we calculate the network's modularity to extract the communities. Gephi does this through a Louvian algorithm \cite{Blondel2008FastUO} with a resolution term. We keep the resolution parameter at the default of one as this gives a manageable number of communities each of which appear thematically clear. Each of these communities can have over a thousand books in them, too many to look at by hand. We instead look at the most representative books in each community. 

But how do we decide which books are the most representative? Two natural measures are weighted degree and eigenvector centrality. The weighted degree measures the overlap of that books readers/fans with those of other books. A book with high weighted degree is a book that is shared amongst readers of many other books in the genre. The alternative measure is eigenvector centrality. Books which are highly connected to other highly connected books tend to have high eigenvector centrality. A representative book is then one that is centrally located in our similarity network. One interpretation is this: a person just read a book and wants to choose the next book. They choose the next book by choosing a neighboring book in the network with the chance of picking the book proportional to the link weight. If thousands of users did this thousands of times, moving from one book to another with a chance to re-read books, then the eigenvector centrality measures how often a book is read, proportional to all other books.  

Weighted degree and eigenvector centrality can give different results. If we sort by weighted degree and take the top twenty books we often see only half are in the top twenty books when sorted by eigenvector centrality. Because of this we consider both measures when investigating the communities.

Note that when finding the top books in a community we restrict ourselves to only that community during the calculation. Otherwise a book that is in one community but is connected to many books in another community can have high weighted degree or eigenvector centrality without representing its own community. Such books are interesting because they may have higher measures in the full dataset and would represent bridging works which  span communities. Our clustering algorithm would place them in a single community when in reality they might represent two or more. Such situations are beyond the scope of this paper, but are ripe for future research.

Once we have a feel of the communities by studying the top books we take each of the communities and measure the distribution of subjects in them. We then compare the distribution to those of the other communities, and to that of the entire network. If the distribution is sufficiently different then it suggests that subject alone can predict which community a book would be in. And, as subjects are our proxy for traditional genre, it suggests that traditional genre is an effective way to partition books.

The analysis so far has focused on higher-level communities. But sub-communities exist, and the factors that form these may be different. We rerun the analysis above but this time on each community, breaking them down into subcommunities.

Finally, to round the study out, we take a different perspective through a PCA analysis. Here communities are the variables and the subjects are the individuals. This allows us to consider combinations of subjects that best describe our communities. 

%=================================
\section{Results}
%===============================

\begin{figure}[H]
	\centering
	\includegraphics[width=0.9\linewidth]{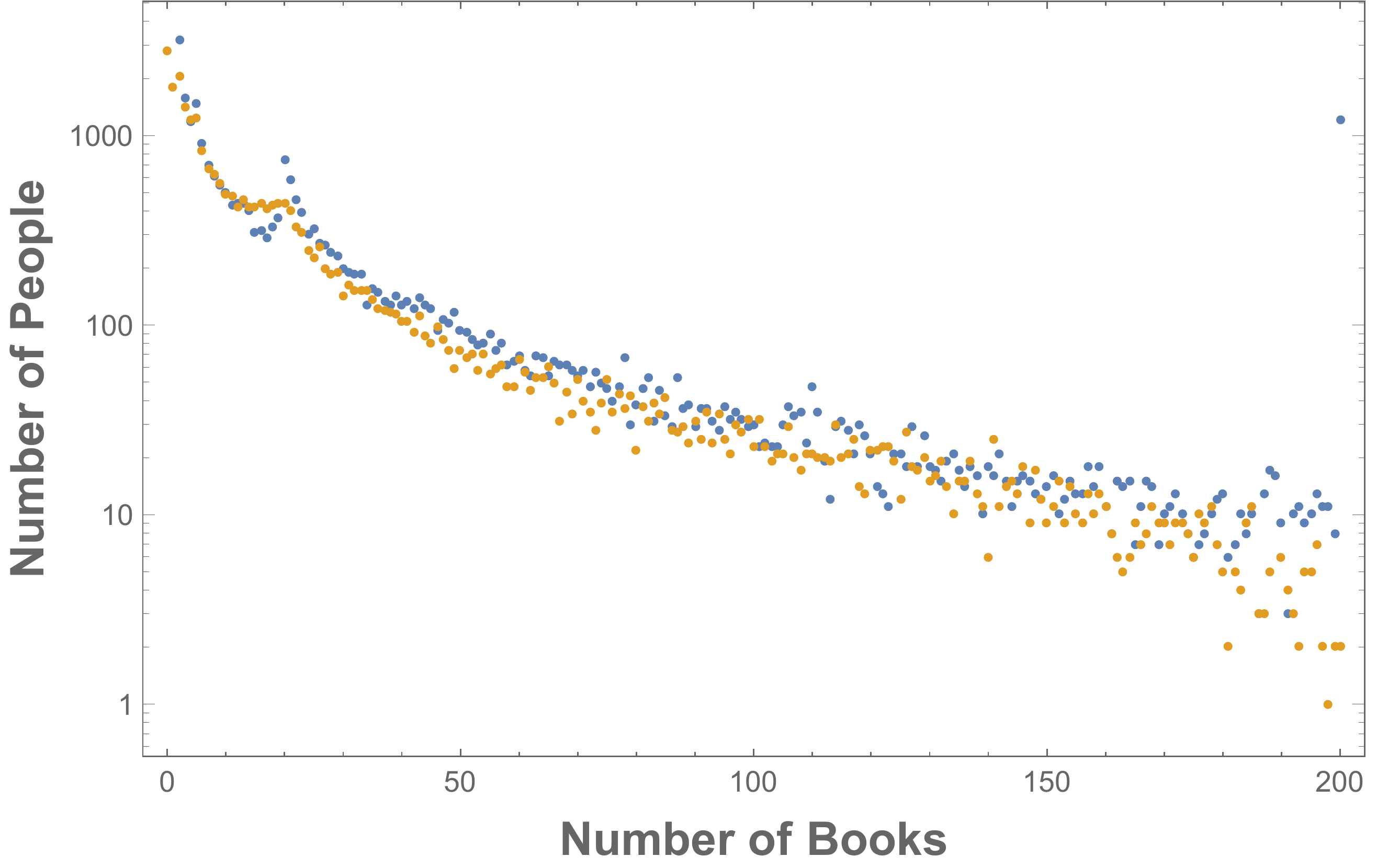}
	\caption{The degree distributions for readers and the number of books they've read (blue) or enjoyed (orange). The last datapoint of the read books suggest that many people have read exactly 200 books. This is an artifact of the Goodreads API, which limits us to 200 books per user. We are unsure what causes the spike at twenty books. It may be the Goodreads reading challenge, which defaults to asking users to read twenty books in a year.}
	\label{fig:number-books}
\end{figure}

\begin{figure}[H]
\begin{center}
\resizebox{.8 \textwidth}{!}{%
	\begin{tabular}{ l  l  l}
    \toprule
	 \textbf{Reader Network} & \textbf{Reader Network} & \textbf{Enjoyment Network} \\
	 \midrule
     Number Books & 10811 & 6770\\
     Number Edges & 6390103 & 2559245\\
	 Average Degree& 1182.15 & 756\\
	 Avg. Weighted Degree & 96.44 &  12.83\\
	 % Network Diameter &  & \\  % This takes a long time to calculate
	 Density & 0.11 & 0.11\\
	 Modularity & 0.302  & 0.487\\
	 Avg. Clustering Coefficient & 0.49 & 0.59\\
	 \bottomrule

	\end{tabular}
	}
	\captionof{table}{A few choice network metrics comparing the two projected networks.  }
	\end{center}
\end{figure}

\begin{figure}[H]
\begin{center}
\resizebox{\textwidth}{!}{%
	\begin{tabular}{ l  l  l  l  l}
    \toprule
	 \textbf{Measure} & \textbf{Reader} & \textbf{Reader (filtered)} & \textbf{Enjoyment} & \textbf{Enjoyment (filtered)} \\
	 \midrule
	 Number Readers & 26076 & 26076 & 26076 & 26076\\
     Number Books & 279152  & 10811 & 279152 & 6770\\
     Number Edges & 943856 & 2559245 & 577650 & 313111\\
     \midrule
      Average Rating &4.06 & 4.09 & 4.56 & 4.58 \\
	 Avg. User Degree & 36.20 & 19.94 & 22.15 & 12.01 \\
	  Avg. Book Degree & 3.38 & 48.10 & 2.07& 46.25 \\
	 \bottomrule

	\end{tabular}
	}
	\captionof{table}{Statistics for the bipartite networks, either with all links (reader) or links with ratings four or more (enjoyment). These networks do not include data from users with less than two books. In the filtered versions of the network we also remove any book with less than ten reads. Notice that this raises the average rating only slightly, even though one might expect that rarely read books are not enjoyable and that is why they are rarely read. The average ratings and other properties of rarely read books are not investigated further in this paper but would be interesting future work.}
	\end{center}
\end{figure}

\subsection{Network Communities}

The reader network and the enjoyment network each give different communities. We name the communities by restricting the network to only that community and then considering the top twenty most representative books for both weighted degree and eigenvector centrality. We also confirmed that the community name remained consistent with less representative books along with the distribution of subject within a community.

%==============================

\begin{figure}[h]
	\begin{center}
		\resizebox{.7\textwidth}{!}{%
		\begin{tabular}{ l l }
			\toprule
			\textbf{Community Name} & Percent of Reader Network  \\
			\midrule
			Young Adult & 28.48\%\\  % Class 4
			Contemporary/Realistic & 26.88\% \\  % Class 3
			(Modern) Classics & 24.66\% \\  % Class 1
			Genre Fiction & 15.09\% \\  % Class 2
			Children's &  4.53	\% \\  % Class 0
			In Death Series & 0.36\%\\  % Class 5
			\bottomrule
		\end{tabular}
	}
	\end{center}
\end{figure}

\begin{center}
	\resizebox{.7 \textwidth}{!}{%
	\begin{tabular}{ l l }
    \toprule
	 \textbf{Community Name} & Percent of Enjoyment Network  \\
	 \midrule
	 (Modern) Classics & 24.96\% \\  % Class 3
	 % modern classics dominate but older ones are lower on the list
	 Contemporary/Realistic & 21.70\% \\  % Class 5
	 Young Adult & 21.33\% \\  % Class 6
	 Fantasy/Scifi & 13.13\% \\  % Class1
	 Children's &  9.99\% \\  % Class 4
	Thriller &  7.47\% \\  % Class 0
	% Looks like weighted genre for this class is dominated by a few super series.
	 Manga &  1.42\% \\  % Class 2
	 \bottomrule
	\end{tabular}
}
	\captionof{table}{The reader and enjoyment communities along with the percentage of the network they make up. Because our books consist of only the well read books on Goodreads the percentages may not reflect the true total percentage of books. (For example, there may be swaths of generic fantasy which nobody reads and so does not appear in our data). Thriller and Fantasy/Scifi from the enjoyment network are merged into Genre Fiction within the reader network. Likewise Manga is a subset of Young Adult in the reader network.}
\end{center}

%The relative sizes of the communities do not contain much information because of how the networks change size when we filter away unenjoyed books. If some communities are less popular then we are less likely to sample positive reviews from them and this could bias the change in size.

  %These communities are mostly equally big, apart from childern's. If we change the community finding algorithm to make more communities then parts of the large communities start to break off into small sub-communities. For example, changing the algorithm to make six communities causes young adult to form a vampire-romance sub-community.
%
%\begin{center}
%
%    \begin{tabularx}{\linewidth}{| X | X | X | X | X | }
%    \textbf{Scifi and Fantasy} & \textbf{School Reading} & {Young Adult} & {General Fictions + Memoir} & {Thriller and Horror}\\
%    \hline
%    Lord of the Things & The Alchemist & Hunger Games & The Help & The Da Vinci Code
%    \end{tabularx}
%
%
%	\captionof{table}{Top ten books in each community (by weighted degree).}
%\end{center}
%

Almost all these communities seem reasonable. But strangely the In Death series forms its own community in the reader network. In Death is a fifty six book detective series written by Nora Roberts. Each book in this series is standalone and can be read independently. We suspect it forms its own community because of the size of the series and the tendency of readers read books within a series. 

Sometimes when we run the clustering algorithm the number of communities can shift by one or two as the algorithm is stochastic. The runs used for this paper are breakdowns we felt representative of a typical run. Still, we briefly investigated how community structure changes as we change the number of communities. When increasing the number of reader network communities to equal the number of enjoyment network communities we found the Realistic/Contemporary community splits into two parts, though the difference between them isn't clear except one may contain more thriller-style books. When we reduce the number of enjoyment network communities to equal the number of reader network communities then Manga, Young-adult, and (Modern) Classics all combine into a single community, which we hypothesize may consist mainly of books read in high school.

Note that we lack a community for purely non-fiction works. Even when we instruct the clustering algorithm to make more communities it does not create a clear non-fiction community. We suspect this is because few non-fiction works exist in our data-set. While Goodreads lists many non-fiction books it has few users that rate those books. This is likely because Goodreads is targeted primarily as a fiction platform, and either attracts readers that prefer fiction or causes the users to not add non-fiction books to their account. When nonfiction does appear it is generally in the (Modern) Classics community.

See the supplementary materials for tables of the top books in each community along with how they are broken down into subcommunities.

\begin{figure}[H]
	\centering
	\includegraphics[width=1\linewidth]{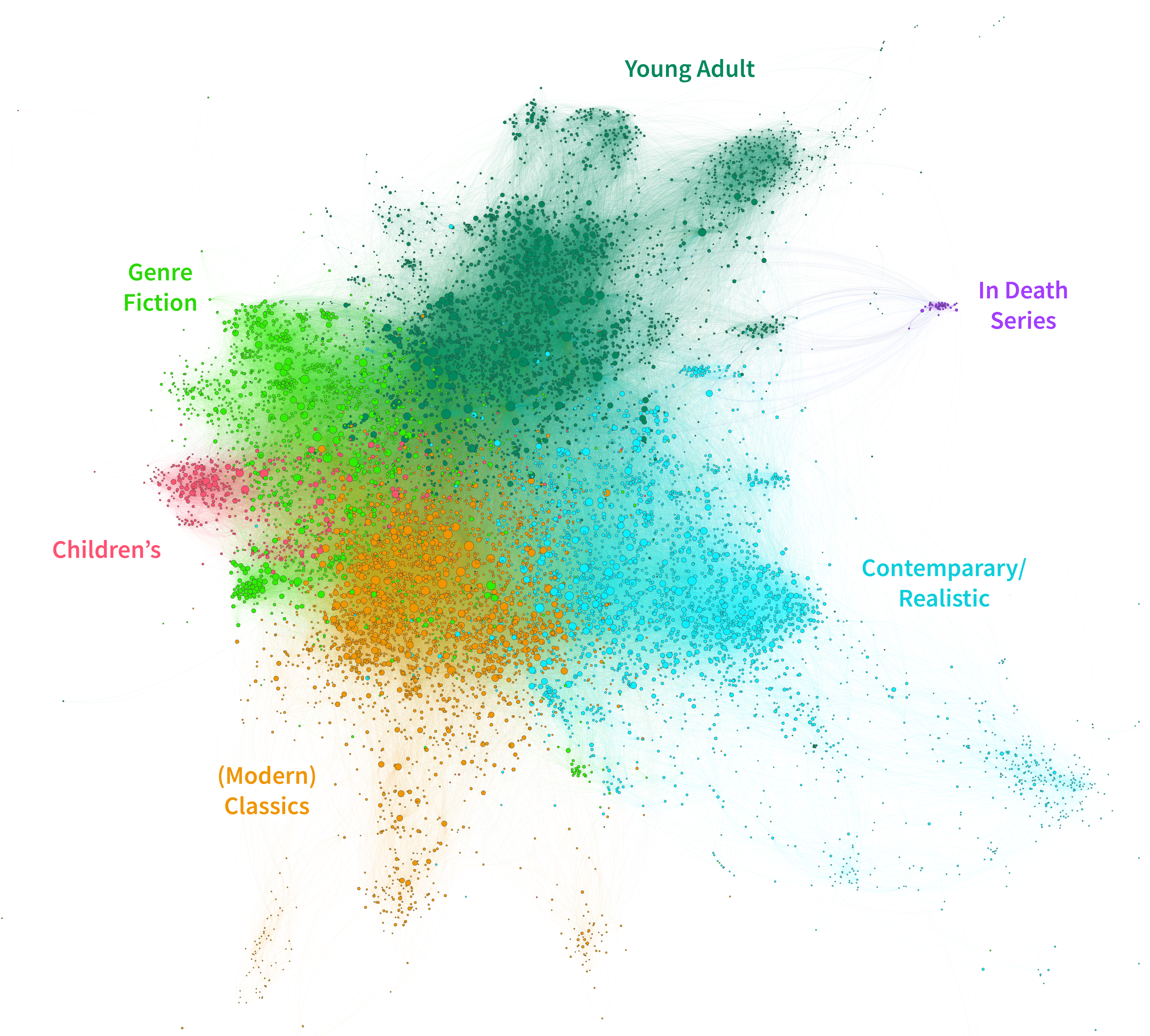}
	\caption{The reader network, with reader genres labeled. An explorable version of this network is \textit{Reader Network Full.gephi} within the supplementary materials. }
	\label{fig:networksreader}
\end{figure}

\begin{figure}[H]
	\centering
	\includegraphics[width=1\linewidth]{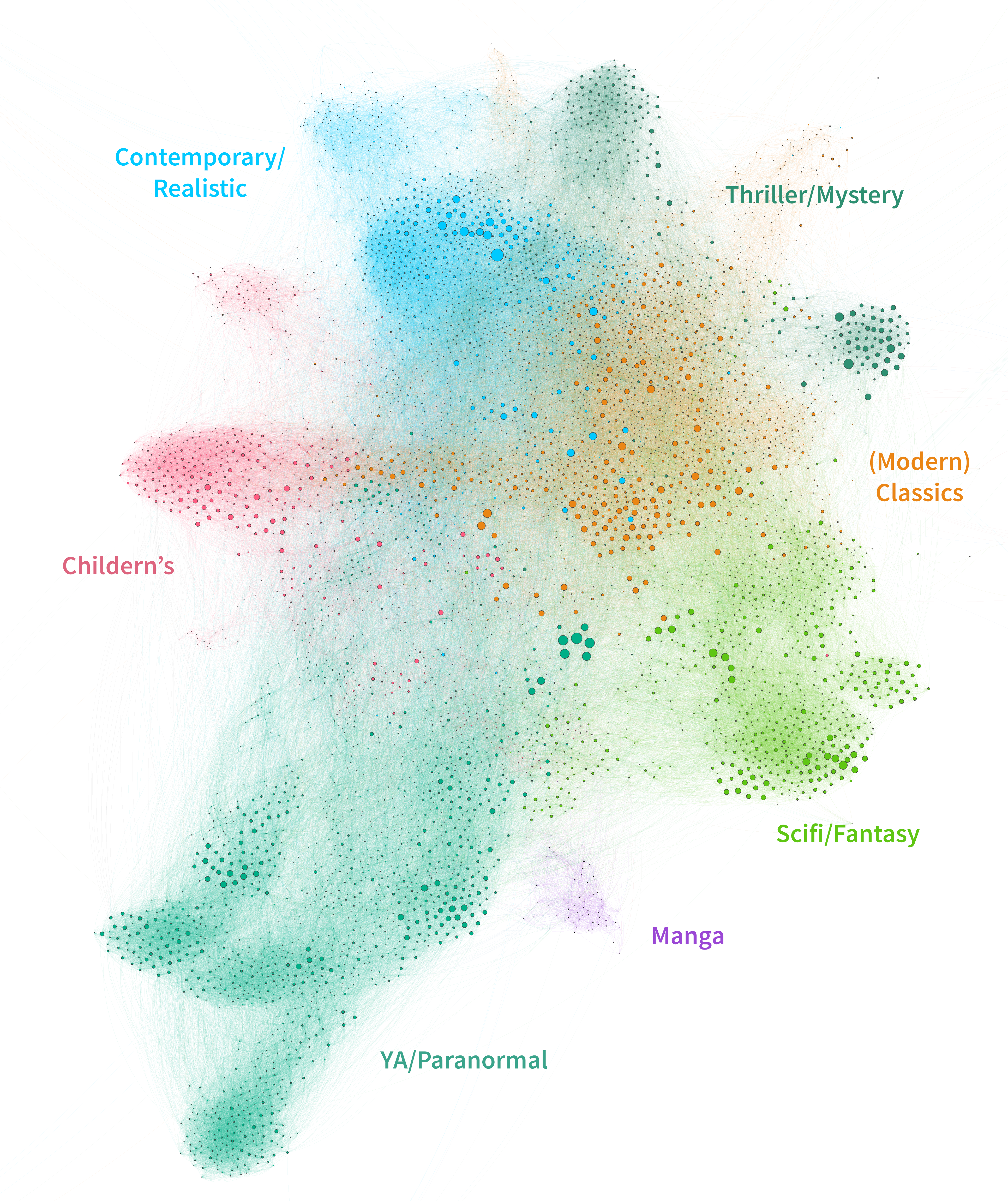}
	\caption{The enjoyment network, with enjoyment genres labeled.An explorable version of this network is \textit{Enjoyment Network Full.gephi} within the supplementary materials.}
	\label{fig:networksenjoyment}
\end{figure}

\subsection{PCA Analysis and Axes of Repulsion}

The PCA analysis for the reader network gives us five total axes, four of which explain significant proportions of the variance.

In the enjoyment biplot we see that Fantasy/Scifi/Young Adult and arguably Manga all tend towards one direction. This is opposite to more realistic fiction. We also find that Thriller and Children's are almost orthogonal to each other. %realism/fantasy axis and a adult/children's axis.

% Reader Table
\begin{table}[H]
	\centering
	\resizebox{\textwidth}{!}{%
		\begin{tabular}{@{}llll@{}}
			\toprule
			\multicolumn{2}{c}{Dimension   1 (33\%)}            & \multicolumn{2}{c}{Dimension 2 (26\%)}                  \\  \cmidrule(r){1-2}  \cmidrule(r){3-4}
			Positive Extreme   & Negative Extreme        & Positive Extreme           & Negative Extreme    \\ \midrule
			Criticism And Interpretation & Discworld (Imaginary Place)           & Criticism And Interpretation          & Horror Fiction                        \\
			English Literature & Wizards                 & English Literature         & Thrillers           \\
			Textual Criticism  & Demonology              & Textual Criticism          & Suspense            \\
			English Drama                & High Schools                          & English Drama                         & Stephanie Plum \\
			Poetry             & Werewolves              & Specimens                  & Maine               \\
			Russia             & Hugo Award Winner       & Toy And Movable Books      & Suspense Fiction    \\
			Business           & Award:Hugo\_Award=Novel & Classic Literature         & Horror              \\
			Description And Travel       & Princesses                            & Children's Stories, English           & New Jersey                            \\
			India              & Boarding Schools        & Folklore                   & Horror Tales        \\
			Success            & Imaginary Places        & History And Criticism      & Suspense            \\
			&                         &                            &                     \\ \toprule
			\multicolumn{2}{c}{Dimension 3 (23\%)}              & \multicolumn{2}{c}{Dimension 4 (18\%)}                  \\ \cmidrule(r){1-2}  \cmidrule(r){3-4}
			Positive Extreme   & Negative Extreme        & Positive Extreme           & Negative Extreme    \\ \midrule
			Horror Fiction               & Stephanie Plum (Fictitious Character) & Discworld (Imaginary Place)           & Werewolves                            \\
			Horror Tales       & Historical              & Hugo Award Winner          & Psychic Ability     \\
			Horror             & Roman                   & award:hugo\_award=novel    & Boarding Schools    \\
			Thrillers          & New Jersey              & Life On Other Planets      & Greek Mythology     \\
			Maine              & Secrecy                 & Imaginary Wars And Battles & High Schools        \\
			Sherlock Holmes    & Fiction / Literary      & Wizards                    & Horror Fiction      \\
			Science Fiction              & Family Secrets                        & Stephanie Plum  & School Stories                        \\
			Serial Murders     & Best Friends            & Imaginary Places           & Young Adult Fiction \\
			Boston             & Life Change Events      & English Fantasy Fiction    & Spy Stories         \\
			Suspense           & New York (State)        & Chicago (Ill.)             & Horror             \\  \bottomrule
		\end{tabular}%
	}
	\caption{The four important dimensions of the PCA analysis, along with the most extreme subjects within them. The percentage next to each dimension gives the proportion of the variance they explain. These can be thought of axes in which the subjects at each extreme repulse subjects from the other extreme. For example, in the first dimension English Literature opposes Wizards, meaning if somebody reads books about wizards they are much less likely to read English literary works. }
	\label{tab:PCA-reader-dim}
\end{table}

\begin{figure}[H]
	\centering
	\includegraphics[width=1\linewidth]{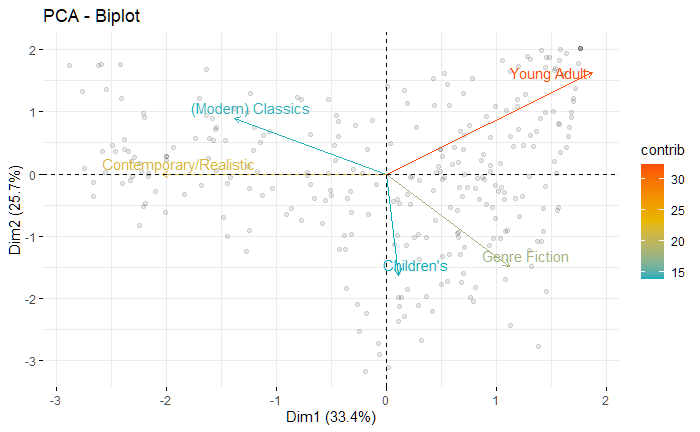}
	\includegraphics[width=1\linewidth]{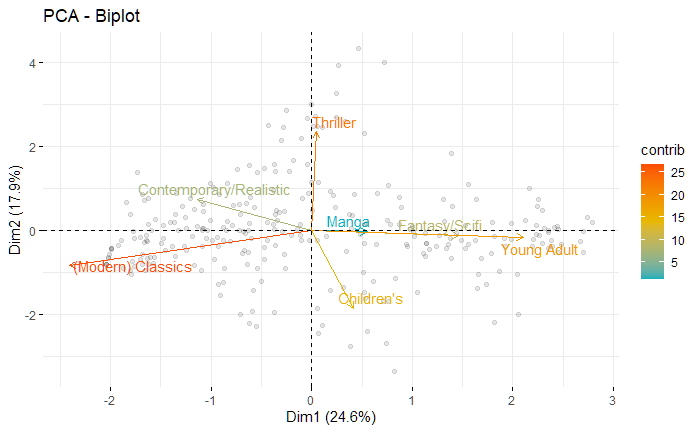}
	\caption{Where each community are located relative to the first two PCA dimensions for the reader network (top) and enjoyment network (bottom). We call the first dimension (y-axis) the maturity axis and the second the realism axis. Together this form a maturity-realism plane.}
	\label{fig:pca-enjoyment}
\end{figure}

\begin{figure}[H]
	\centering
	\includegraphics[width=0.7\linewidth]{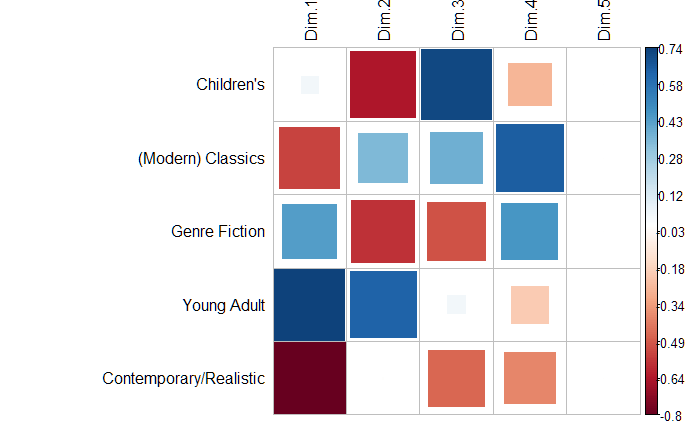}
	\includegraphics[width=0.7\linewidth]{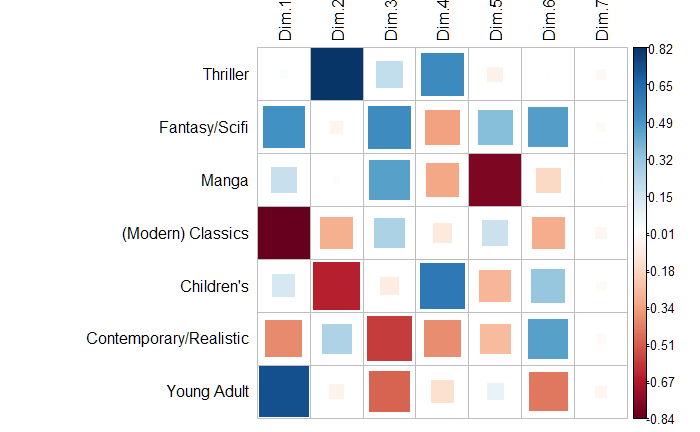}

	\caption{The corresponding correlation plots for reader (top) and enjoyment (bottom) communities.}
	\label{fig:pca-corr}
\end{figure}

 \section{Discussion}

We investigated how well traditional genres describe what books people read and enjoy. Communities of books in both the reader network and enjoyment network turn out to be combinations of traditional genres, mimicking findings in music genres classification \cite{lambiotteUncoveringCollectiveListening2005a}. This suggests that traditional genres are already a powerful system of classifying books and, more generally, stories. 

Interestingly there are differences in what people read and what they enjoy. Enjoyment appears more specific than reading, as evidenced by the more detailed breakdown of our communities. Furthermore, our PCA analysis showed that a large amount of variance between our communities could be described in terms of two axes: one which measures realism and one which measures maturity.   

\subsection{Revisiting Elemental Genres}

What do these results say about elemental-genres? While each community can be described by traditional genres this does not mean elemental-genre wrong or useless. Our study cannot measure the emotional content of books but we can speculate with the data we do have.

If the core of a story is indeed its emotional thrust we still may not see elemental-genres appear in the data is because the established structures of traditional genre overpower it. People may tend to stay in a single traditional genre simply because it is the best system available. 

This may explain the reader data but it is harder to believe for enjoyment. One imagines the enjoyment reflects the true feelings of a the reader and is less influenced by the traditional genre which labels the story. The caveat here is that the traditional genre of a story may change the reader's perception of it. People may rate certain books highly to project a type of persona (remember that Goodreads ratings are public and shared with a user's friends). For example, classic literature may receive higher ratings than a typical thriller to signal a reader is sophisticated.

Another possibility is that elemental-genre is deeply interwoven with traditional genre. After all, traditional genre captures tone along with setting, plot, and tropes. Certain elemental-genres may correlate with certain traditional genres, intertwining to the point where our analysis cannot separate them. If this is true then considering both traditional and elemental genre points to unexplored combinations: a thriller with the emotional beats of contemporary/realistic fiction, or a fantasy/scifi with the emotions of classic literature.

Returning to this idea that people read to ``hack'' their brain into feeling certain emotions, it is possible that this is true but at different moments of life people want to experience different emotions. Readers may then bounce around elemental-genres but still prefer a certain traditional genre, so end up reading within their traditional genre. In such a world the data we would show communities that represent traditional genres, but be made up of many elemental genres.

Alternatively, traditional genre and elemental-genre are both forces, but traditional genre is the more influential of these. In this case traditional genre might appear as the higher level communities but the subcommunities will be made up of elemental-genres. Subcommunities may map onto elemental-genres (see supplementary materials). A full analysis of this possibility is beyond our scope and requires subject experts in each of the communities or subcommunities.

We must keep in mind throughout all of this that traditional genre evolve and adapt to the current needs of readers, booksellers, and writers. As new unexplored combinations of genres appear and gain popularity booksellers will create a new genre for those books. If a pure elemental-genre became popular it would certainly find itself one day called traditional genre. One could say this has already happened to many of the elemental-genres given in table \ref{tab:elemental-genres}. Because of this effect we may never be able to cleanly tease apart elemental-genre and traditional genre; the terms are too fuzzy and leak into each other over time.

The evolution of traditional genre leads to interesting questions. \cite{sackSimulatingCulturalEvolution2013} showed through a cultural simulations that traditional genres may change over time with old ones vanishing and new ones appearing, and that this is synced with the generation time of readers. If this is true could elemental-genre be a constant, appearing in different forms but always present? If so it deserves the name``elemental." Or does it too change over time? Could we delve into literary history and discover forgotten emotional-resonances? Would these tell us about the emotions people of that different age wished to feel?

The above arguments detail ways which elemental-genre could coexist with our results that traditional genre is a good classification system. But we have the last possibility: elemental-genre is a negligible or redundant, and any importance it holds is already twined together with traditional genre. The status-quo holds and elemental genre is at best a mental framework for writers. 

\subsection{What people read versus what people enjoy}

The communities representing what one is likely to read and what one is likely to enjoy differ in specificity. This can be seen by how the reader network community of Genre Fiction is splits into two communities in the enjoyment network: Thriller and Fantasy/Scifi. Likewise the reader network community of Young Adult is made up of Young Adult and Manga. The latter we did not expect to find as a high level genre due to its lower relative readership in the west and on Goodreads. Perhaps fans enjoy the combination of pictures and the Japanese story telling structure.  It is also surprising that Romance does not appear as its own community in either of the networks. It is rather split into Young Adult Romance and Realistic/Contemporary Romance. %\needcitation. \todo{Look where other Japanese books appear in this network}

Recommendation engines must take care to decide whether to classify in terms of readership or enjoyment. This might take the form of a trade-off of deciding if one wants a higher probability of a sale now (if so classify by reader network communities) or a high rate of satisfaction (if so classify by enjoyment network communities). The latter increases the probability of a good review which generate more sales \cite{chevalierEffectWordMouth2006a}.

\subsection{PCA analysis}

Our PCA analysis gives axes which correspond to combinations of traditional genres. These can be considered a spectrum reflecting the emotional tone of the story. The PCA biplot for the enjoyment network is easiest to interpret so we focus on those. Somewhat similar interpretations hold for the reader network, but the results are not as clean.

The first dimension (y-axis) appears to reflect reader age. In the positive coordinates we find Stephen King and other adult thriller and horror books while in the negative coordinates we find Dr. Seuss and other children's books. Between these we see Young Adult and the other communities that could be read by either children or adults. We call this axis the maturity dimension. Mature books will likely have a much gritter and darker tone than non-mature books.

The second axis (x-axis) we call a realism dimension. Realistic/Contemporary and (Modern) Classics tend to sit on the realism side of this axis while Young Adult and Fantasy/Scifi tend to sit on the fantasy side. Manga and Children's also tend towards the fantasy side while Thriller remains firmly in the middle.

If these interpretations are correct then the two features that most define a story is its maturity level (ie. what age group it is aimed at) and how realistic it is. These aspects form a spectrum, and the two together form a two dimensional plane on which we can place a story. 

This ``maturity-realism" plane may be a useful framework as a coarse way to label types of stories. It can also point to holes in the market. The first quadrant is mostly empty. This suggests there are fewer subjects, and therefore books, which are both mature and fantastical. By considering the other axes we can further hone into these holes. These are potentially areas to avoid because the combination does not work. Alternatively, they may be fertile ground for untapped markets. It has been claimed that it is the new combinations of existing genres have birthed some of our most successful books. Twilight is a combination of paranormal and romance. Harry Potter is a combination of boarding schools and fantasy. Dune is a science-fiction with the world-building depth of an epic fantasy.

% A full analysis of this sort would first need to plot books relative to the PCA axes. Then one would look for areas that are empty. These are either untapped niches to fill or difficult areas that nobody has yet had success in. To find the latter areas plot successful and unsuccessful books and see areas where only unsuccessful books appear.

% Of note is that there are blank spots in this biplot. In the enjoyment biplot we see that the first quadrant is empty compared to the others. This might suggest new niches available to fill. In this case this would be a Fantasy/Thriller combination. 
%The fact the axes are opposed that does not mean they cannot be mixed  (As it happens, there already exists a literary magazine devoted toliterary fantasy: Unedr Ceaseless Skys)

\subsection{Alternative classification systems as a tool}

Traditional, elemental, reader, enjoyment -- multiple classification systems like these form tools for those who work in the fiction industry. The obvious one is recommendation systems. Instead of giving a user a list of recommendations in the typical collaborative filtering recommendation style one could instead switch between these alternative systems. This may be preferable to giving a list of top recommendations because many users enjoy the ability to browse. An explorable network visualisation, with the ability to switch between similarity measures, may quickly lead readers to their next book. We provide such networks in our supplementary materials in the \textit{Enjoyment Network Full} and \textit{Reader Network Full} Gephi files.

Data about the enjoyment-communities and the maturity-realism plane may also be of great use to writers and editors. Fans of an established author expect the author's new book to have all the elements they love of that author while being different and distinct enough to not feel like a repeat of previous works. Authors must balance the new and the familiar. A typical recommendation is to pick up books within the traditional genre they wish to write, find out what the audience enjoys about them, and make sure their book has many of those elements while subverting a few of them. This same advice is also often given to new authors who wish to establish themselves.  \cite{cardElementsFictionWriting2010}. 

Our study provides a complement to this. An author can first get a coarse sense of the location of their audience on the maturity-realism plane. Then for finer details the author can look at enjoyment network communities that their book could fall in. If they already have published books they can see what other books they are linked to. This allows the author a systematic sampling of books the beyond a single traditional genre.

Writing a novel is a large time investment, and consecutive book deals are often based on their previous books’ performance. It is reasonable for authors to be  conservative, staying with stories they know work instead of branching out. The enjoyment network along with the maturity-realism plane are tools that can allow an author to experiment while reducing the chances of their work being unsuccessful. %Publishers can also take into account the reader network to estimate how well a book would sell.

\subsection{Future Directions}

Much research remains. We have ignored the bridges between communities. Considering the edges between two communities would give us a sense of how similar the communities are and allow us to see which books are highly connected to two or more communities. If done carefully this could shed light on the claim that books which mix genres tend to be the breakout successes. These mixed community books in principle could combine the familiar and unfamiliar while attracting readers from both communities, giving that book a large audience. 

Likewise the maturity-realism plane we develop hints at mature-fantasy being an underdeveloped genre. Success of books like Game of Thrones suggest that this combination is rising in popularity. One could study the other PCA axes and search for more combinations that are rare.

One factor we did not consider in detail is the author. Our data suggests that author recognition likely plays a large role, as expected \cite{yucesoySuccessBooksBig2018a} \cite{yuPracticalTypologyAdult1999}. Because large series can easily rise to the top in the data from their sheer number we suggest a future analysis creates an author-similarity network rather than a book-similarity network. Authors often write in one style, taking up pen names when changing genre. This will give us another, maybe better, perspective on these community breakdowns.

A caveat with our approach is that there is no consensus on what traditional genre a book resides in. Star Wars is the typical example of this. Does it belong in the traditional genre of fantasy or science fiction? Should it instead belong in some separate specific genre of science-fiction-fantasy or a general overarching genre of speculative fiction? By allowing books to have multiple subjects we avoided this problem. Yet we enforced that the network communities are create are distinct, with no overlap. Sub-communities we build are similarly separated. Our analysis allows no blurred boundaries even though in reality genres seep into each other. This is a technical limitation that could be fixed in future studies.

%Because of this we take two approaches. The first is to pick a "canonical" source of traditional genres. This data is difficult to get.

%We relied upon the ---------- community detection algorithm to identify genres. This can create a hierarchy of genres, where each sub-genre is nested in larger genres but no genres at the same level in the hierarchy can overlap.

%The means that in our framework Star Wars would be able to reside in the high-level-genre of speculative fiction, the middle-level genre of either fantasy or science fiction, and a low-level subgenre of science-fiction-fantasy. However, it could not be in both the middle level genres of fantasy and science fiction.

%An open problem is to generalize our framework to the overlapping genre case.
%Implementing this should not be too hard, as there already exists algorithms for weighted overlapping bipartite clustering. \needcitation \todo{If it is so easy then why don't we do it ourselves? Maybe should.}

%---------------------

\subsection{Big data as a tool for authors}

Our goal is not only to study reader and enjoyment genres but to also give examples of how big data give rise to new fiction writing techniques. There exists a vast literature on fiction writing techniques. These range from overarching principles of story structure to specific tips on stylizing prose and dialogue. These can be found in many writing book and generally come by way of applied psychology and accumulated wisdom of previous writers \cite{cardElementsFictionWriting2010, cron2012wired, maass2009fire, vogler2007writer, le2015steering, frey2010write}.  Yet more books are being published than ever before. We have data, and by careful analysis of this data we can create new tools for fiction writers so that fiction is no longer only an art, but also a science. 

We've considered reader-communities and enjoyment-communities as ways to craft books readers will both choose to read and enjoy. We've also shown the potential of our maturity-realism plane to coarsely identify books one might enjoy, and more generally the potential of PCA analysis to find untapped combinations of genres. 

One can easily imagine an array of other data inspired approaches. Natural language processing represents a major area here. For example, it is known that given a book the number of nodes and the average shortest path length of the book's word network is correlated with the style of certain literary eras \cite{Amancio_2012}. If an author wishes to mimic the style of an era but their prose doesn't feel quite right they could look at the word network their book generates. Perhaps they find that the size of their word network is too small, meaning they are using too few unique words. This can help ``debug" prose.

Adapting this technique could help character sound distinct. A common problem is that different character's dialogue sound the same simply because the same author wrote it. If one can extract the each of the main character's dialogue (or the prose of their point of view scenes) then one can run those texts through text-author identification or natural language processing methods. The data from these can warn when characters sound the same and how to fix it. 

For a final example an author could write a few lines then ask a neural network to continue it. What the neural network gives often has no plot but are ripe with interesting turns of phrases and directions to take a scene. This allows the author to inject into their process a little outside creativity and chaos.

None of these techniques replace writers. They do not swap art with science. Instead they complement a skilled writer's intuition. A century ago we built the typewriter to increase a writer's speed. Decades ago we built the word processor, granting authors quick and easy restructuring of text. The data driven ideas we build here -- book networks, communities, and maturity-realism planes -- are examples of the same: tools for the making of art. 

\subsection*{Acknowledgments}

Many thanks to John O'Donovan, Ambuj Signh, Tim Robinson, and many peers of the UC Santa Barbara Network Science IGERT for helpful discussion and suggestions that guided the early stages of this research. Rachel Redberg helped code and conceptualize an early pilot version of this project.

\printbibliography
\end{document}